\documentclass[iop]{emulateapj}
\usepackage{graphicx}    % needed for including graphics e.g. EPS, PS
\usepackage{color}
\submitted{}

\newcommand{\Msun}{M_{\odot}}

\newcommand{\Mvir}{M_{\rm vir}}

\newcommand{\Mgas}{M_{\rm gas}}
\newcommand{\Mstar}{M_{\rm star}}

\newcommand{\Mcs}{M_{\rm 1}}
\newcommand{\Mch}{M_{\rm 2}}

\newcommand{\MvirB}{\mathbf M_{\rm {\bf vir}}}
\newcommand{\MstarB}{\mathbf M_{\rm {\bf star}}}
\newcommand{\MgasB}{\mathbf M_{\rm {\bf gas}}}

\newcommand{\MvirMstarB}{  $\MstarB${\bf (}$\MvirB${\bf )}   }
\newcommand{\MstarMvirB}{  $\MvirB${\bf (}$\MstarB${\bf )}   }
\newcommand{\MgasMstarB}{  $\MgasB${\bf (}$\MstarB${\bf )}   }
\newcommand{\MstarMgasB}{  $\MstarB${\bf (}$\MgasB${\bf )}   }

\begin{document}
\title{GalMass: A Smartphone Application for Estimating Galaxy Masses}
\author{Kyle R. Stewart\altaffilmark{1,*}}
\affiliation{Jet Propulsion Laboratory, California Institute of Technology, Pasadena, CA 91109, USA}
\altaffiltext{1}{NASA Postdoctoral Program Fellow}
\altaffiltext{*}{For GalMass related comments and inquiries please email galmass@krstewart.com.
An IDL package that duplicates the various conversions outlined here 
is available at $\texttt{http://www.krstewart.com/galmass/galmass.pro}$.}

\begin{abstract} {This note documents the methods used by the smartphone application, ``GalMass,''
which has been released on the Android Market.  GalMass estimates the
halo virial mass ($\Mvir$), stellar mass ($\Mstar$), gas mass ($\Mgas$), and galaxy gas fraction of a central galaxy
as a function of redshift ($z<2$), with any one of the above masses as an input parameter.
In order to convert between $\Mvir$ and $\Mstar$ (in either direction), GalMass uses
fitting functions that approximate the abundance matching models of
either \cite{cw09}, \cite{Moster10}, or \cite{Behroozi10}.  GalMass uses a a semi-empirical fit to
observed galaxy gas fractions to convert between $\Mstar$ and $\Mgas$,
as outlined in \cite{Stewart09b}.
}

\end{abstract}

\section{Introduction}
\label{Introduction}
In the LCDM model of structure formation, dark matter (DM) halos form by the hierarchical accretion
of smaller dark matter halos, as well as more diffuse accretion of matter that is not
virialized \citep[e.g.][]{Stewart08}.
In this model, galaxies reside at the center of dark matter halos, such that the mass
scale of a galaxy's environment can be characterized by the virial mass of the dark matter halo
in which it resides.  This mass scale influences a great number of galaxy properties, including
luminosity, morphology, star formation rate, gas content, and color.  As such, theoretical
formalisms often categorizes dark matter halo properties as a function of virial mass.

Unfortunately, halo virial masses are difficult to determine, empirically.
Instead, observations are more likely to categorize galaxies by their stellar mass (or luminosity).
As a result, it is often difficult to compare theoretical expectations with observations
because the mapping between stellar mass and virial mass is non-trivial, and evolves over
time \citep[e.g.,][]{Behroozi10}.  For example, theoretical measures of the dark matter halo merger
rate typically describes mergers in terms of the ratio of dark matter masses prior to the merger,
while observational studies often refer to the \emph{stellar} mass ratio,
even though a $1:3$ ratio in dark matter can easily correspond to a $1:20$ or $1:2$ ratio in stellar
mass, depending on the mass regime in question \citep{Stewart09conf}.
Even in a less formal setting than scientific articles,
theorists and observers often ``think'' in terms of these different quantities, making comprehensive
discourse of galaxy evolution difficult.
The goal of GalMass is to convert between different galaxy masses
quickly and accurately from the convenience of a smartphone application.

A secondary goal of GalMass is to encourage the scientific community to
consider the development of scientific-level applications in the growing
smartphone marketplace (e.g., the ``Android Market'' or the Apple's
``App Store'').  While a number of astronomy applications exist in these
markets, they are typically intended for a public audience.  There are
very few applications that are developed from within the scientific community
for productive use in the scientific community (examples of
such scientific applications developed by professional astronomers
include GravLensHD, CosmoCalc, and Exoplanet)\footnote{These applications are available
on the iPhone.}.

\section{Abundance Matching Models}
\label{stars}
One method for mapping between the dark matter virial mass ($\Mvir$) and the corresponding stellar mass ($\Mstar$)
presumes a monotonic relationship between halo mass and stellar mass.
That is, so long as a galaxy is the central galaxy to its dark matter halo (not a satellite galaxy),
more massive galaxies are presumed to reside in more massive DM halos.
In this model, one can achieve a mapping between halo mass and stellar mass by
setting the abundance of halos more massive than $\Mvir$, $n_h(>M_{\rm DM})$, equal to
the abundance of galaxies more massive than $\Mstar$, $n_g(>M_{\rm star})$.  (Note that in order to correctly account
for satellite galaxies using this method, one must track each satellite
back to when it was last isolated, matching the abundance of its DM halo at that time,
instead of using its current subhalo mass.)  This technique,
often referred to as ``abundance matching'' has been shown to reproduce a number of galaxy
clustering statistics, as well as matching observed close pair counts of galaxies
as a function of redshift \citep[e.g.,][]{conroy06,Berrier06,Shankar06,Stewart09a}.

While a monotonic relationship between halo mass and stellar mass cannot be
true for every galaxy on a case-by-case basis, especially at lower masses where galaxy formation is expected
to be more stochastic \citep[e.g.,][]{BoylanKolchin11}, in a statistically averaged sense it provides
a good characterization of the relationship between halo mass and stellar mass that must be true
in order for LCDM halos to reproduce the observed stellar mass function.
Indeed, even when extrapolating abundance matching
relations to very low halo masses where it is not well-tested, it has been shown to produce
a plausible (if not robust) model for the dwarf galaxy satellite population around
Milky Way size halos \citep{stealth}.

In order for GalMass to quickly and accurately estimate virial and stellar masses using this
abundance matching formalism, it utilizes analytic fitting functions as approximations to
published relations.  The curve fitting routine used in constructing these fits is the publicly
available ``pymodelfit'' routine\footnote{This python package, created and maintained by
Erik Tollerud, was also used to create the best-fit ``fundamental curve'' of
\cite{Tollerud11}.  It is freely available at $\texttt{http://packages.python.org/PyModelFit/index.html}$},
which minimizes $\chi^2$ in order to determine the best-fit
parameters of a custom parametrization to a set of data.
Currently, GalMass allows for $3$ choices for the abundance matching model used: that of
\cite{cw09}, \cite{Moster10}, and \cite{Behroozi10}, hereafter CW09, M10, and B10 respectively.

\subsection{Fitting Functions: \citealp{cw09}}
As \cite{cw09} includes no analytic functional forms to is abundance matching results,
I choose the following functional form to characterize $\Mstar(\Mvir)$:
\begin{equation}
\label{bigeq}
\Mstar = \Mcs \frac{\Mvir^{\alpha} }{\Mch^{\beta} } \left(\frac{\Mch + \Mvir}{2}\right)^{(\beta-\alpha)}
\end{equation}
where $\alpha$ and $\beta$ are the low and high mass slopes, and the transition
between these power laws is described by a characteristic halo mass, $\Mch$
(with a normalization corresponding to a characteristic stellar mass $\Mcs$).  Minimizing $\chi^2$ to determine the
best-fit parameters, I find this functional form to be a good approximation for the results of CW09\footnote{See \S\ref{AMdiscussion}
for discussion on the goodness of fit of the equations presented here.}.
Once a fit has been found at each epoch ($z<2$), I find that the redshift dependence of each parameter
(or the log of parameters $\Mcs$ and $\Mch$)
is well-characterized by either a linear, quadratic, or power law relation in $z$, as seen in
the right column of Table \ref{tablefits}.

Because Equation \ref{bigeq} is not easily reversible, I adopt the same functional form
in order to independently determine an analytic fit for $\Mvir(\Mstar)$:
\begin{equation}
\label{bigeq2}
\Mvir = \Mcs \frac{\Mstar^{\alpha} }{\Mch^{\beta} } \left(\frac{\Mch + \Mstar}{2}\right)^{(\beta-\alpha)}
\end{equation}
with $\Mcs$ now representing
a characteristic \emph{stellar} mass, and $\Mch$ giving the normalization (halo mass).
Again, I find this form to provide a good fit to the CW09 data at each epoch, with the redshift evolution
of each parameter given by a quadratic in $z$.  These fits are shown in the left column of Table \ref{tablefits}.

\subsection{Fitting Functions: \citealp{Moster10}}
For M10, a fitting function for $\Mstar(\Mvir)$ is already provided, including redshift evolution.
For the sake of completeness, I include this fitting form (their Equation 2) here:
\begin{equation}
\label{M10big}
\Mstar= 2\Mvir \left(\frac{m}{\Mvir}\right)_0 \left[ \left(\frac{\Mvir}{M_1}\right)^{-\beta} +  \left(\frac{\Mvir}{M_1}\right)^{\gamma}\right]^{-1}
\end{equation}
where $(m/\Mvir)_0$ provides the normalization at the characteristic halo mass $M_1$ and $\beta$ and $\gamma$ control
 the behavior at the low and high mass ends.  The redshift dependence of their free parameters are
 given in M10 as power laws (or linear relations) in $1+z$.  These equations (from their Equations 23-26 and
 Table 7) as well as the equation above are included in the right column of Table \ref{tablefits}.

Again, this fitting form is not easily reversible.  In order to obtain an analytic fit for
$\Mvir(\Mstar)$, I use the above fit from M10 to construct a discrete set of
data points with which to fit Equation \ref{bigeq2}.
This functional form provides a good parametrization to M10's results,
with the redshift evolution of each parameter given by the left column of Table \ref{tablefits}.

\subsection{Fitting Functions: \citealp{Behroozi10}}
B10 gives an analytic fit for $\Mvir(\Mstar)$, but not vice versa
(opposite from that of M10, which provides a fit for $\Mstar(\Mvir)$).
For simplicity's sake, I adopt their fitting function for the case with no presumed systematic errors (their $\mu=k=0$ case).
Their fitting form for $\Mvir(\Mstar)$ (their Equation 21) is as follows:
\begin{equation}
\label{B10big}
\begin{array} {lcl} \log_{\rm 10}(\Mvir) & = & \log_{\rm 10}(M_{\rm 1}) + \beta \log_{\rm 10}(\Mstar/M_{\rm 0}) \\
                                         &  & + \frac{(\Mstar/M_{\rm 0})^{\delta}}{1+(\Mstar/M_{\rm 0})^{-\gamma}} -\frac{1}{2} \end{array}
\end{equation}
where $M_{\rm 1}$ and $M_{\rm 0}$ are the normalization and characteristic mass, $\beta$ controls the
low mass end slope, and $\delta$ and $\gamma$ both control the high mass slope.  The redshift evolution of
the various parameters are given by linear relations in $(1-a)$, where $a$ is the scale factor,
defined by $a=1/(1+z)$ such that a value of $a=1$ is present day.  The various redshift evolutions
of these parameters (from their Equation 22 and Table 2) are all included in the left column of
Table \ref{tablefits}.

As with M10, this fitting form is not trivially reversible, and I use their fit to construct a
discrete data set, from which I fit Equation \ref{bigeq}, in order to parameterize $\Mstar(\Mvir)$.
The fitting function and redshift evolution of each parameter is given in the right column of
Table \ref{tablefits}.

%------------------------------------- Galaxy Properties TABLE ----------------------------
\begin{table*}[p!]
\begin{center}
\renewcommand{\arraystretch}{1.5}
\renewcommand{\tabcolsep}{.005\textwidth}
\caption{FITTING FUNCTIONS USED BY GALMASS}
\label{tablefits}
\begin{tabular}{ | l | l | l |}
  \hline
  & & \\
  {\bf \cite{cw09}}     & \hspace{5em} \MstarMvirB$\tablenotemark{$\ddagger$}$ & \hspace{5em} \MvirMstarB$\tablenotemark{$\ddagger$}$\\ \hline
  Fitting Function      \rule{0cm}{0.6cm} & $\Mvir=\Mcs \Mstar^{\alpha} \Mch^{-\beta} \left(\frac{\Mch + \Mstar}{2}\right)^{(\beta-\alpha)}$
                        & $\Mstar=\Mcs \Mvir^{\alpha} \Mch^{-\beta} \left(\frac{\Mch + \Mvir}{2}\right)^{(\beta-\alpha)}$ \\ [1.8ex] \hline
  Normalization         & $\log_{\rm 10}(\Mcs) = 0.352z^2 -0.285z +12.359$ &$\log_{\rm 10}(\Mcs)  = 0.056z^2 + .068z + 9.5$     \\ \hline
  Characteristic Mass   & $\log_{\rm 10}(\Mch) = 0.102z^2 -0.089z + 10.740$ &$\log_{\rm 10}(\Mch)  = 0.320z^2 + .018z +11.2$     \\ \hline
  Low Mass Slope        & $\alpha = -0.0183z^2 + 0.01726z + 0.3238$        &  $\alpha = .021z^{4.86} + 3.39$       \\ \hline
  High Mass Slope       & $\beta  = 0.1214z^2 -0.9053z + 3.170$            &  $\beta  = .085z + .36$        \\ \hline
  \hline

  & & \\
  {\bf \cite{Moster10}} & \hspace{5em} \MstarMvirB$\tablenotemark{$\ddagger$}$ & \hspace{5em} \MvirMstarB$\tablenotemark{$\dagger$}$ \\ \hline
  Fitting Function      \rule{0cm}{0.6cm} & $\Mvir=\Mcs \Mstar^{\alpha} \Mch^{-\beta} \left(\frac{\Mch + \Mstar}{2}\right)^{(\beta-\alpha)}$
                        & $\Mstar = 2 \Mvir \left(\frac{m}{\Mvir}\right)_0 \left[ \left(\frac{\Mvir}{M_1}\right)^{-\beta} +
                          \left(\frac{\Mvir}{M_1}\right)^{\gamma}\right]^{-1}$ \\ [1.8ex] \hline
  Normalization         & $\log_{\rm 10}(\Mcs)   = 0.022z^2 -0.0185z + 12.544$ &  $(m/\Mvir)_0= 0.0282 (1+z)^{-0.72}$        \\ \hline
  Characteristic Mass   & $\log_{\rm 10}(\Mch)   = 0.0321z^2 -0.138z + 10.846$  &  $\log_{\rm 10}(\Mcs)  = 11.884 (1+z)^{0.019}$ \\ \hline
  Low Mass Slope        & $\alpha = 0.473z -0.0330$                 &  $\beta= 0.17z + 1.06$       \\ \hline
  High Mass Slope       & $\beta  = -0.588z^{0.489} + 2.64$         &  $\gamma= 0.556 (1+z)^{-0.26}$        \\ \hline
  \hline

  & & \\
  {\bf \cite{Behroozi10}} & \hspace{5em} \MstarMvirB$\tablenotemark{$\dagger$}$ & \hspace{5em} \MvirMstarB$\tablenotemark{$\ddagger$}$\\ \hline
  Fitting Function      \rule{0cm}{0.6cm} & $\log_{\rm 10}(\Mvir)=\log_{\rm 10}(M_{\rm 1}) + \beta \log_{\rm 10}(\Mstar/M_{\rm 0}) $
                        & $\Mstar=\Mcs \Mvir^{\alpha} \Mch^{-\beta} \left(\frac{\Mch + \Mvir}{2}\right)^{(\beta-\alpha)}$ \\
                        & $\hspace{15ex}+ \frac{(\Mstar/M_{\rm 0})^{\delta}}{1+(\Mstar/M_{\rm 0})^{-\gamma}} -\frac{1}{2} $ &   \\ [1.8ex] \hline
  Normalization         & $\log_{\rm 10}(M_{\rm 1}) = 12.35 - 0.30 (1-a)$\tablenotemark{$\star$} &
                          $\log_{\rm 10}(\Mcs) = 0.0350 z^2 -0.192 z + 10.199$ \\ \hline
  Characteristic Mass   & $\log_{\rm 10}(M_{\rm 0}) = 10.72 - 0.59 (1-a)$\tablenotemark{$\star$} &
                          $\log_{\rm 10}(\Mch) = 0.00509 z^2 +0.00299 z + 11.824$  \\ \hline
  Low Mass Slope        & $\beta = 0.43 - 0.18(1-a)$\tablenotemark{$\star$}  &  $\alpha = -0.2076 z^2 + 0.752 z + 2.423$       \\ \hline
  High Mass Slope       & $\delta  = 0.56 - 0.18(1-a)$\tablenotemark{$\star$}    &  $\beta  = 0.120 z^2 -0.0994 z + 0.206$        \\
                        & $\gamma = 1.54 - 2.52(1-a)$\tablenotemark{$\star$} &  \\ \hline

   & & \\
   {\bf \cite{Stewart09b}}   & \hspace{5em}  \MgasMstarB$\tablenotemark{$\dagger$}$  & \hspace{5em} \MstarMgasB$\tablenotemark{$\dagger$}$ \\ \hline
    Fitting Function    \rule{0cm}{0.6cm} &  $\Mgas = 0.04 \, \Mstar \left( \frac{\Mstar}{4.5\times10^{11}\Msun}\right)^{-\mu}$
                        &   $\Mstar= 25 \, \Mgas \left(\frac{\Mgas}{1.8\times10^{10}\Msun}\right)^{\mu/(1-\mu)}$ \\ [1.8ex] \hline
  Redshift dependence   &  $\mu=0.59 (1+z)^{0.45}$                &   $\mu=0.59 (1+z)^{0.45}$   \\ \hline
  Scatter               \rule{0cm}{0.5cm} & $\sigma(\log_{\rm 10}(\Mgas/\Mstar)) = 0.34 - 0.19 \log (1+z)$
                        & $\sigma(\log_{\rm 10}(\Mstar/\Mgas)) = 0.34 - 0.19 \log (1+z)$   \\ [1.8ex] \hline
  \end{tabular}
  \end{center}
  \tablenotetext{0}{\tablenotemark{$\dagger$}These fits are ``original'' fits given by the corresponding paper.}
  \tablenotetext{0}{\tablenotemark{$\ddagger$}These fits are newly derived from the original data (or fitting functions).}
  \tablenotetext{0}{\tablenotemark{$\star$}$a$ is the scale factor, with $a=1$ being present day, and $a=1/(1+z)$.}
\end{table*}

%------------------------------------- Galaxy Properties TABLE ----------------------------

\subsection{Abundance Matching Discussion}
\label{AMdiscussion}
Because in these models I have independently fit models for $\Mvir(\Mstar)$ and
$\Mstar(\Mvir)$ and there is some margin of error to each
fitting form, I caution that these equations are not entirely self-consistent.  One may use the above fitting
functions to convert from halo mass to stellar mass back to halo mass and not return the
precise beginning value\footnote{I originally designed GalMass
to recursively call $\Mstar(\Mvir)$ when converting from stellar mass to halo mass,
until it produced self-consistent results with $\Mvir(\Mstar)$.  For reasons that are
unclear, this approach was found to significantly slow down the operation, with
lag times of $3-5$ seconds on a typical Android phone.  In the end, I opted for speed over precision.}.

In general, I find that converting from halo mass to
stellar mass using the above fitting functions tends to be more accurate to the original data
of CW09.  (Thus, whenever GalMass shows a small discrepancy between derived values for the CW09 model,
priority should be given to $\Mstar(\Mvir)$, when the input parameter is halo mass).
Similarly, for M10 the original function for $\Mstar(\Mvir$) will be more accurate, by definition.
Since the original fitting function from B10 goes in the opposite direction, priority should
be given to $\Mstar(\Mvir)$ for their model.

Despite these small deviations, the errors introduced by fitting the various models to the functional
forms provided are no larger than the typical uncertainties inherent in the abundance matching technique
\citep[$\sim0.25$ dex; see][]{Behroozi10}.
While I refer the reader to B10 for a thorough exploration of the statistical uncertainties and
errors associated with abundance matching models, I do caution that all of these models are inherently
less certain for $z>1$, due to uncertainties in the observed stellar mass function at higher
redshifts.

\section{Galaxy Gas Model}
\label{gas}
In order to estimate the gas content of galaxies as a function of stellar mass, I
adopt the relation from \cite{Stewart09b}, henceforth S09, which
quantifies observationally-inferred relations between galaxy gas fraction
and stellar mass.  Specifically, S09 characterizes gas fraction data from
\citealp{McGaugh05} (disk-dominated galaxies at $z=0$)
and \citealp{Erb06} (UV-selected galaxies at $z\sim2$) with a
relatively simple function of stellar mass and redshift (their Equation
1).
\begin{equation}
\label{S09a}
\Mgas = 0.04 \, \Mstar \left( \frac{\Mstar}{4.5\times10^{11}\Msun}\right)^{-\mu(z)}
\end{equation}
As with the abundance matching fits, this relation also depends on redshift.  Though
very high stellar mass galaxies tend to have similar gas content at $z=0-2$,
the slope of this relation is much steeper at higher redshift,
a consequence of galaxies being more gas-rich at earlier times.  The redshift evolution of $\mu(z)$
is given in the left column of Table \ref{tablefits}.
Unlike abundance matching, the relation between stellar mass and gas mass
is considerably more stochastic, with significant intrinsic scatter (especially at
low redshift).  While the above equation captures the mean relation, S09 also characterizes
the typical scatter about $\log(\Mstar/\Mgas)$,
characterized (in log space) by a gaussian with standard deviation $\sigma$).
This log scatter is given in the left column of Table \ref{tablefits}.

Because the relation between $\Mstar$ and $\Mgas$ is
significantly less complex than that between $\Mstar$ and $\Mvir$, it is a simple exercise to
invert Equation \ref{S09a} in order to obtain $\Mstar(\Mgas)$.  Similarly, because the scatter
is characterized by the log of the ratio $\Mstar/\Mgas$, the same equation for the magnitude of scatter
is also applicable to the log of the inverse ratio, $\Mgas/\Mstar$ (see
the right column of Table \ref{tablefits}).

\subsection{Gas Model Discussion}

S09 finds that the above characterization, which is defined by the gaseous properties of disk galaxies
at $z=0$ and $z=2$ \citep{McGaugh05,Erb06} is also remarkably consistent with other $z=0$
measurements \citep[e.g.,][]{Kannappan04,Wei10}, as well as intermediate redshift observations
\citep[e.g.,][]{Wright09, Hammer09b}.
GalMass provides two methods for utilizing this galaxy gas model.  The first option is to only
return the \emph{average} gas mass for a disk galaxy of a given stellar mass.  The second option,
labeled ``allow scatter'' uses a pseudo-random number generator to include scatter
(up to $\pm\sigma$) from the mean value.

Note that this gas model is based on low redshift samples of isolated disk galaxies.  It does not claim
to reproduce realistic gas content for satellite galaxies that may have had gas stripped
from their halos upon infall, nor is it designed to reproduce the gas content of elliptical
galaxies.  Specifically, the model is designed to fit observations in the stellar mass range
$10^{8.7} \Msun < \Mstar < 10^{11.5}\Msun$.  While GalMass will allow for input
parameters outside of this range (extrapolating the fitting function to higher or lower
masses), these resulting values will not be robust.

As with the abundance matching data, this relation is also less reliable at
higher redshift, as the data from \cite{Erb06} is based on indirect estimates of galaxy gas
fractions. The maximum redshift at which GalMass will produce outputs based on this gas model is $z=2$.

%>>>>>>>>>>>>>>>>>>>>>>>>>>>>>>>>>corotation vs time<<<<<<<<<<<<<<<<<<<<<<<<<<<<<<<<<
\begin{figure}[t!]
 \includegraphics[width=0.48\textwidth]{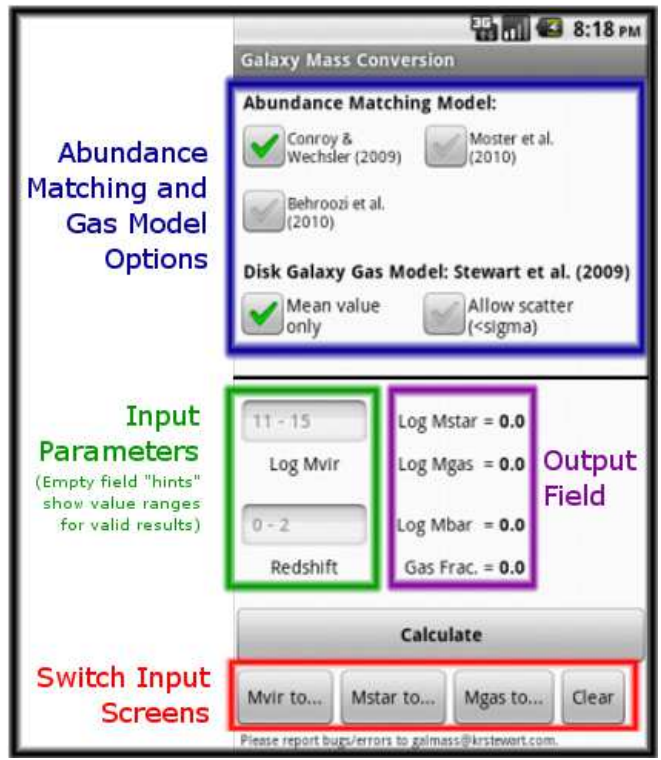}
 \caption{Screenshot of the GalMass application.  The upper (blue) region shows the
 abundance matching and galaxy gas model options (see \S\ref{stars} and
 \S\ref{gas}).  The left (green) and right (purple) regions show the input and output
 fields, respectively.  When the input boxes are empty, they display ``hint'' values
 that span typical mass and redshift ranges where the models are applicable.
 The ``Calculate'' button reads from the input boxes, applies the proper fitting functions,
 and updates the output field.  The bottom (red) region allows the user to switch screens,
 changing the input mass parameter from between $\Mvir$, $\Mstar$, and $\Mgas$.
 The bottom right ``Clear'' button clears all screens
 of their input and output values.
   }
\label{ss}
\end{figure}
%>>>>>>>>>>>>>>>>>>>>>>>>>>>>>>>>>>>>>>>><<<<<<<<<<<<<<<<<<<<<<<<<<<<<<<<<

\section{Implementation}
Throughout the application, masses are always given in units of solar mass,
with log values displayed for compactness.
Halo virial mass, galaxy stellar mass, galaxy gas mass, and galaxy baryonic mass (gas and stars)
are denoted ``Mvir,'' ``Mstar,'' ``Mgas,'' and ``Mbar'' respectively.
The application also computes the galaxy gas fraction (``Gas Frac.'') defined as
$\Mgas/(\Mgas+\Mstar)$.

The basic layout of GalMass consists of three interchangeable screens, depending on the input
parameter desired by the user.  The buttons that switch between screens are found at the bottom
of each screen.  For example, the ``Mvir to...'' button corresponds to the
screen where the halo virial mass is expected as an input, and the galaxy's stellar and gas masses are
given as outputs (see Figure \ref{ss}).

For a more detailed discussion of what mass regimes and redshifts abundance matching results
 can be trusted, I refer the reader to the original papers from whence I have derived fitting
 functions: CW09, M10 and B10.  Still, to aid the user in knowing when
 the calculations are based on extrapolations beyond well-tested regimes,
 GalMass displays a warning message (but output fields are still populated) if any calculations
 result in one the following: $\Mvir>10^{15}\Msun$;
 $\Mvir<10^{10}\Msun$; or $\Mstar<10^{8}\Msun$.  A more severe warning message is displayed
 (and output fields are cleared to null values) if any input or output values contain the following:
 $\Mvir>10^{16}\Msun$; $\Mstar>10^{13}\Msun$; $z>2$; or any mass $<1\Msun$.

 Lastly, I re-emphasize that both the abundance matching data and the galaxy gas models, as implemented
 here in GalMass, are designed for isolated galaxies.  The gas model will almost certainly over-estimate
 the gas content of satellite galaxies, and abundance matching models typically match stellar masses to
 dark matter virial masses upon first infall (when the galaxy was last isolated).

\acknowledgements I would like to thank Charlie Conroy for sharing his abundance matching data from CW09
and Erik Tollerud for help in using his ``pymodelfit'' code.
I would also like to thank Leonidas Moustakas for encouraging me
to consider the possible scientific uses of smartphone applications.
GalMass was constructed with the aid of Google App Inventor (http://appinventor.googlelabs.com/about/).
KRS is supported by an appointment to the NASA Postdoctoral Program at the Jet Propulsion Laboratory,
administered by Oak Ridge Associated Universities through a contract with NASA.
Copyright 2011 California Institute of Technology. Government sponsorship acknowledged.  All rights reserved.

\bibliography{galmass}

\end{document}